\documentclass[12pt,a4paper,final]{iopart}

\usepackage{iopams}  
\expandafter\let\csname equation*\endcsname\relax
\expandafter\let\csname endequation*\endcsname\relax
\usepackage{amsmath}
\usepackage{amssymb}
\usepackage[font=small]{caption}
\usepackage{graphicx}
\usepackage{cite}
\usepackage[breaklinks=true,colorlinks=true,linkcolor=blue,urlcolor=blue,citecolor=blue]{hyperref}

\newcommand{\be}{\begin{equation}}
\newcommand{\ee}{\end{equation}}
\newcommand{\bal}{\begin{aligned}}
\newcommand{\eal}{\end{aligned}}

\begin{document}

\title[Lorentzian Spectral Geometry with Causal Sets]{Towards Spectral Geometry for Causal Sets}

\author{Yasaman K. Yazdi$^{1,2}$ and Achim Kempf$^{1,2,3,4}$}
\address{$^1$Dept. of Physics, University of Waterloo, Waterloo, ON, N2L 3G1, Canada}
\address{$^2$Perimeter Institute for Theoretical Physics, Waterloo, ON, N2L 2Y5, Canada}
\address{$^3$Dept. of Appl. Mathematics, Univ. of Waterloo, Waterloo, ON, N2L 3G1, Canada}
\address{$^4$Inst. for Quantum Computing, Univ. of Waterloo, Waterloo, ON, N2L 3G1, Canada}

\ead{yyazdi@pitp.ca, akempf@pitp.ca}

\begin{abstract}
We show that the Feynman propagator (or the d'Alembertian) of a causal set contains the complete information about the causal set. Intuitively, this is because the Feynman propagator, being a correlator that decays with  distance, provides a measure for the invariant distance between pairs of events.
Further, we show that even the spectra alone (of the self-adjoint and anti-self-adjoint parts) of the propagator(s) and d'Alembertian already carry large amounts of geometric information about their causal set.  This geometric information is basis independent and also gauge invariant in the sense that it is relabeling invariant (which is analogue to diffeomorphism invariance). We provide numerical evidence that the associated spectral distance between causal sets can serve as a measure for the geometric similarity between causal sets.  
\end{abstract}


\section{Introduction}

The construction of the theory of quantum gravity will involve the construction of a mathematical framework that naturally combines the mathematical frameworks of general relativity and quantum theory, i.e., that combines differential geometry and functional analysis.
The differential geometry of a spacetime is known to consist of the causal structure (i.e., the light cones) together with a conformal volume function \cite{hawk,mal,hawk2}. Causal set theory uses this fact to provide a natural ultraviolet-regularized description of spacetime that is  covariant \cite{cs}. 

In the present paper, we consider the causal set description of spacetimes and we ask to what extent this geometric information can be expressed in functional analytic terms such as the spectra of operators.

Our starting point is the observation that the Feynman propagator expresses the strength of the correlations between quantum field fluctuations at pairs of points. Since these correlations drop with the invariant distance of the two points, the Feynman propagator in effect provides a measure for the invariant distance between points in spacetime. The Feynman propagator could therefore substitute for the metric. Correspondingly, instead of using rulers and clocks (which in any case do not exist at extremely small scales) it is possible, in principle, to measure distances in spacetime by measuring the correlations of quantum fluctuations of fields \cite{msa}.

In this paper, we show for $2$-dimensional spacetimes described by causal sets that the Feynman propagator (as well as the d'Alembert operator) does indeed contain all metric information: knowing the Feynman propagator is to know the causal set. The Feynman propagator on causal sets therefore provides, in this sense, a quantitative measure of the invariant distances between events of its causal set. 

Further, we consider the spectral geometry of causal sets. Traditional spectral geometry \cite{gilk,milnor,sg1}, 
asks how much geometric information about a compact Riemannian manifold is contained in the spectra of Laplacians on that manifold (or also, for example, how much of the shape of a drum one can hear in its spectrum \cite {kac}). See \cite{dh} for a review. The spectral geometry of spacetimes, i.e., of Lorentzian manifolds, however, is still in its infancy.

Here, we consider spacetimes described by causal sets and we calculate the spectra of their correlators and d'Alembert operators, or more accurately, of their self-adjoint and anti self-adjoint parts. We find numerical evidence that these spectra contain a large amount of geometric information: It occurs relatively rarely that, for example, the d'Alembertian spectra of two distinct causal sets coincide.  Indeed, we find numerical evidence that, in general, the more geometrically different two causal sets are, the more their spectra differ. This means that the spectral distances of causal sets could serve as a measure of their geometric similarity. 

We begin with a brief introduction to causal set theory in Section 2, followed by an introduction to quantum field theory on causal sets in Section 3. In Section 4 we present the result that in the two-dimensional theory of a scalar field on a  causal  set, the causal set can  be fully  determined  by  the  Feynman propagator of the theory. In Section 5 we explore the spectral properties of some operators on causal sets for the set of all 6 element and 7 element causal sets. Finally, in Section 6 we study the spectral distances of causal sets sprinkled into different spacetimes.

\section{Causal Set Theory}

Causal set theory, see \cite{cs}, provides a natural locally Lorentz-invariant framework for discretizing spacetime. It is based on the fact that the metric of a Lorentzian spacetime consists of the causal structure and a conformal scaling function. Namely, 
causal set theory  is a theory of discrete spacetime elements  or `atoms' and the causal relations amongst them. Their density encodes the conformal volume factor. The discrete elements and the ordering relation induced by the causal structure together form a partially ordered set, $\mathcal C$. The set $\mathcal C$ and the ordering relation $\preceq$ satisfy

\begin{itemize}
\item Reflexivity: for all $X\in \mathcal C$, $X \preceq X$.\

\item Antisymmetry: for all $X,Y\in \mathcal C$, $X\preceq Y\preceq X$ implies $X=Y$.\

\item Transitivity: for all $X,Y,Z\in \mathcal C$, $X\preceq Y\preceq Z$
implies $X\preceq Z$.\

\item Local finiteness: for all $X,Y\in \mathcal C$,
$|I(X,Y)|<\infty$, where $|\cdot|$ denotes cardinality and $I(X,Y)$
is the causal interval defined by $I(X,Y):=\{Z\in \mathcal C|X\preceq Z\preceq Y\}$.\\
\end{itemize}
We will write $X\prec Y$ if $X\preceq Y$ and $X\neq Y$.

As we mentioned, this approach is founded on work by Hawking \cite{hawk} and Malament \cite{mal} which shows that the causal structure of a spacetime, together with a conformal factor, determines the metric uniquely. In causal sets, the spacetime volume of a region is given by the number of elements it contains, and this provides the additional information of the conformal factor. Causal set theory is unique in that it discretizes spacetime while  preserving local Lorentz invariance. 

\begin{figure}[htb!]
 \centering
 \includegraphics[width=.65\textwidth]{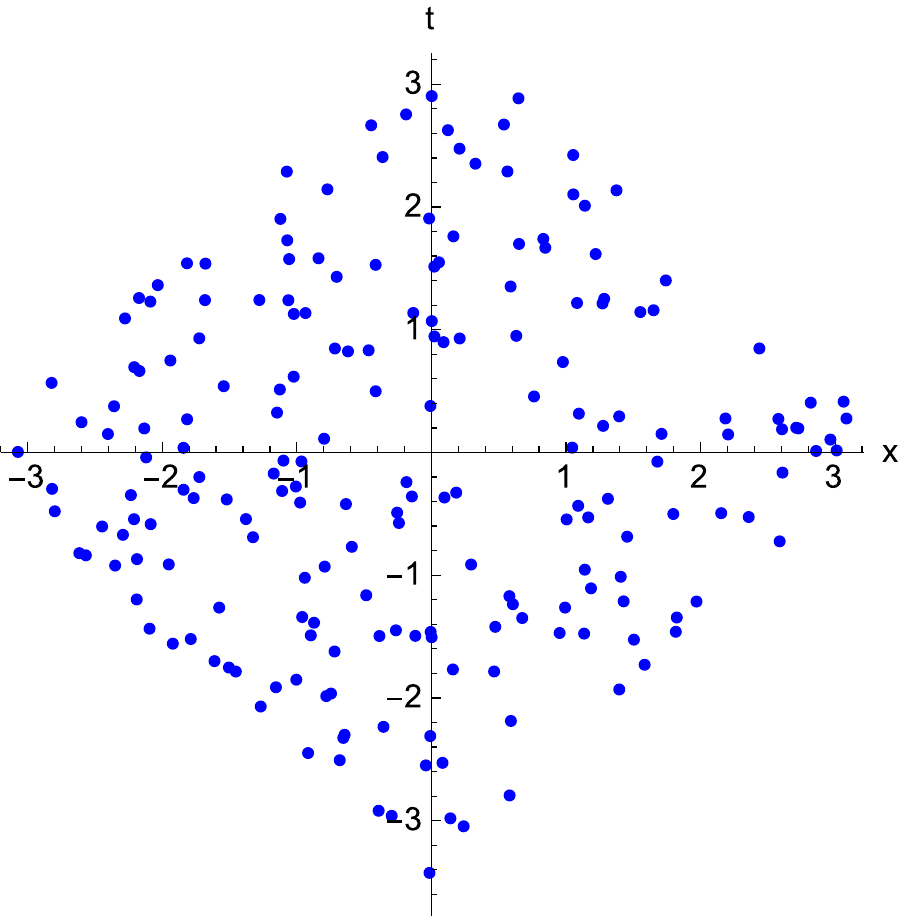}
 \caption{\label{2dcs}A causal set formed by sprinkling 200 elements into a finite interval in $1+1$ dimensional Minkowski spacetime.}
\end{figure}

In this paper we will primarily consider $2d$ causal sets. A causal set consisting of points sprinkled into a region of $\mathbb{M}^2$ is shown in Figure \ref{2dcs}. 

A useful way to represent a causal set is through its causal matrix $C$ or link matrix $L$. The causal matrix is defined by
\be
C_{xy}=\begin{cases}
1, & \text{for $x\prec ~y$}\\
0 & \text{otherwise}
\end{cases}
\ee
A link (or nearest neighbour relation) is a relation $x \prec y$ such that there exists no $z \in \mathcal{C}$ with $x\prec z\prec y$. We then say that $x$ and $y$ are nearest neighbours $x \prec\!\!* ~y$. The link matrix, $L$,  is defined by
\be
L_{xy}=\begin{cases}
1, & \text{for $x\prec\!\!* ~y$}\\
0 & \text{otherwise}
\end{cases}
\ee

We can always choose a labelling (called a \emph{natural labelling}) to ensure that $C$ and $L$ are both strictly upper triangular.

\section{Quantum Field Theory on a Causal Set}

In this section we outline a set of useful propagators whose expressions have a known form in causal set theory.

The retarded Green's function (in $2d$) is given by \cite{sj1}

\begin{equation}
    G_R:=\frac{1}{2}C\left(I+\frac{m^2}{2\rho}C\right)^{-1}
\label{ret},
\end{equation}

where the constant $\rho$ is the sprinkling density and $m$ is the particle mass. The field commutator $\left[\phi(x),\phi(x')\right]=i\Delta(x-x')$, also called the Pauli-Jordan function, is given by

\begin{equation}
    \Delta:=G_R-G_A,
\end{equation}
where $G_A$ is the advanced Green's function and is equal to the transpose of the retarded Green's function. $i \Delta$ is anti-symmetric and Hermitian. Its non-zero eigenvalues come in pairs of positive and negative real numbers. Restricting to the positive eigenspace of $i \Delta$, we can define a two-point correlation function or Wightman function for the theory
\begin{equation}
    W:=\text{Pos}(i\Delta).
\end{equation}
The choice of vacuum state that this prescription leads to is called the Sorkin-Johnston state \cite{sj1, sj2}.

The Feynman propagator, in terms of the operators we have just defined, is

\begin{equation}
    G_F=G_R+iW.
    \label{gf}
\end{equation}
These functions agree  with their continuum counterparts in the limit of high density (see eg. \cite{sj1, yas1}).

\section{Causal Sets in terms of Scalar Field Propagators}
In this section we extend some of the results of \cite{msa} for continuum theories to the case of $2d$ causal sets. In that work it was shown that the metric tensor can be reconstructed from the inhomogeneous propagators of a scalar quantum field. We ask whether a causal set can be determined from knowledge of only the propagator of the scalar field theory on the causal set. We specifically consider the Feynman propagator defined in the previous section:

\begin{equation}
    G_F=G_R+\text{Pos}\, i(G_R-G_R^\dagger).
    \label{gf2}
\end{equation}
It is clear that given a $G_R$ there is a unique corresponding $G_F$. Is the reverse also true? Given a $G_F$ is there a unique $G_R$ it will correspond to? Let us assume that there exists another retarded Green's function $\tilde{G}_R$ from which $G_F$ could also be constructed:

\begin{equation}
    G_F=\tilde{G}_R+\text{Pos}\, i(\tilde{G}_R-\tilde{G}_R^\dagger).
\end{equation}
$G_R$ and $\tilde{G}_R$ can be made to be strictly upper triangular while $\text{Pos}\, i (G_R-G_R^\dagger)$ and  $\text{Pos}\, i (\tilde{G}_R-\tilde{G}_R^\dagger)$ are Hermitian. Therefore the difference between 
$G_R$ and $\tilde{G}_R$ cannot be compensated for by the difference between $\text{Pos}\, i (G_R-G_R^\dagger)$ and  $\text{Pos}\, i (\tilde{G}_R-\tilde{G}_R^\dagger)$. It follows that $\tilde{G}_R=G_R$ and there is a unique $G_R$ corresponding to each $G_F$. Therefore, knowledge of $G_F$ indeed implies complete knowledge of the causal set.

Operationally, the Feynman propagator $G_F$, being central to the Feynman rules, can in principle be measured through suitable  particle physics experiments. This would mean that we can measure $G_F$, deduce $G_R$ and therefore the causal matrix $C$. Once we have $C$ we know what the causal set is.

\begin{figure}[htb!]
 \centering
 \includegraphics{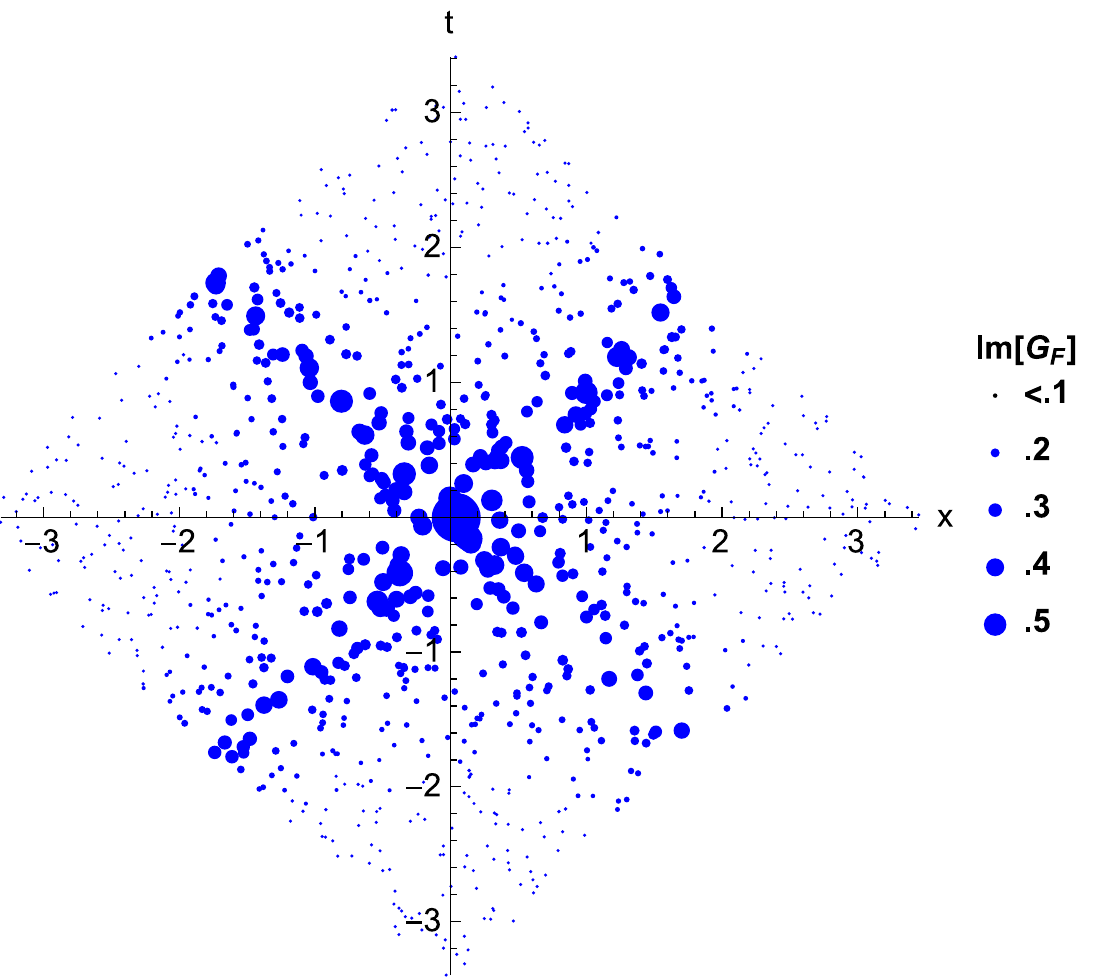}
 \caption{\label{imgf} The imaginary part of the massless Feynman Propagator from an event at the center to other events indicated by dots. The background causal set is a sprinkling of 1000 elements into a finite interval in $1+1$ dimensional Minkowski spacetime, with $x,\, t$ coordinates shown on the axes. The magnitude of its imaginary part, $\text{Im}[G_F]$, is indicated by the radius of the dots. The magnitude of $\text{Im}[G_F]$ decays with the distance away from the lightcone of the point at the center. This shows that the imaginary part of the Feynman propagator contains the information about the amount of invariant distance that there is between two events - except that it does not tell us if this distance is spacelike or timelike.}
\end{figure}

\begin{figure}[htb!]
 \centering
 \includegraphics{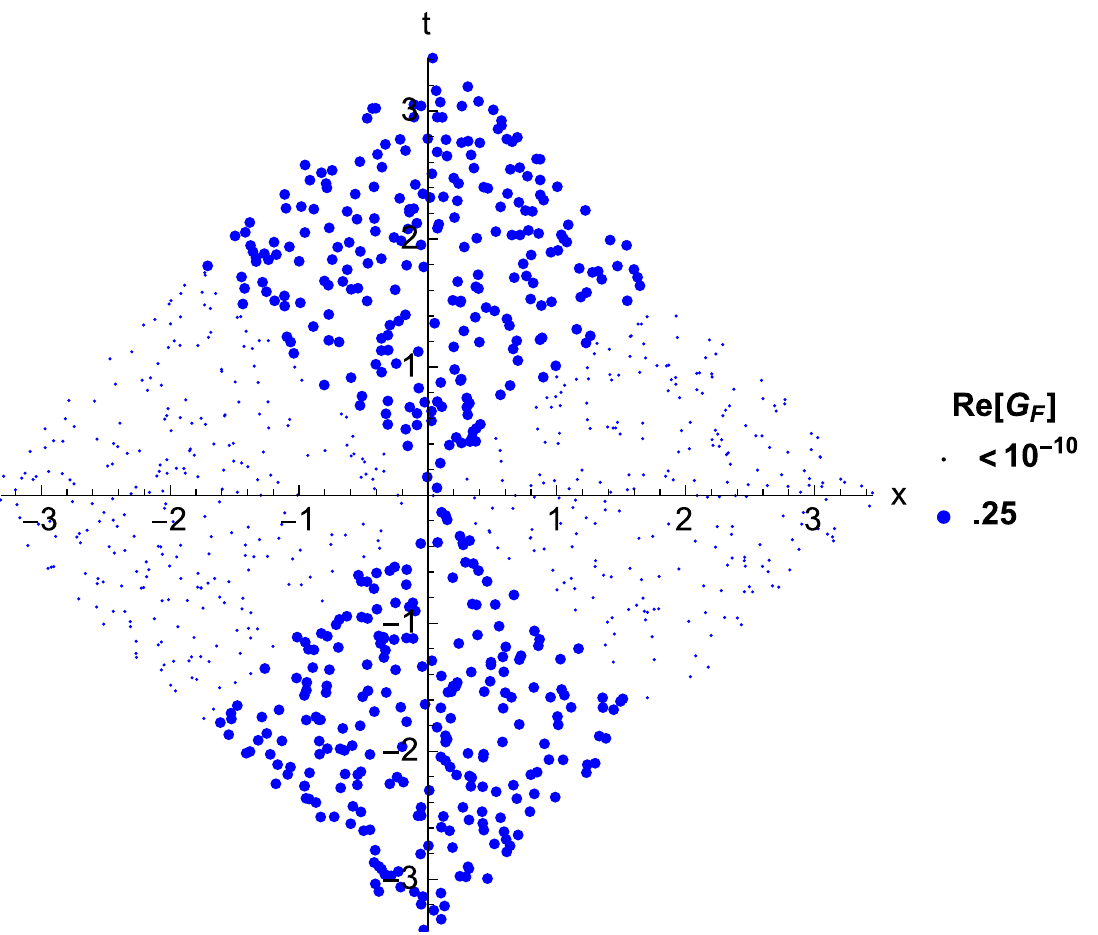}
 \caption{\label{regf}The real part of the massless Feynman propagator from an event near the center to other events indicated by dots. The background causal set is a sprinkling of 1000 elements into a finite interval in $1+1$ dimensional Minkowski spacetime, with $x,\, t$ coordinates shown on the axes. The magnitude of its real part, $\text{Re}[G_F]$, is indicated by the radius of the dots. The magnitudes are $\frac{1}{4}$ inside the lightcone and close to zero outside of the lightcone of the event near the center. This shows that the real part of the Feynman propagator carries the information about whether two events are spacelike or timelike.}
\end{figure}

Figures \ref{imgf} and \ref{regf} show the imaginary and real parts of the Feynman propagator with respect to one point fixed near the center. Larger dots correspond to larger magnitudes for $G_F$. As shown in Figure \ref{imgf}, the magnitude of the imaginary part, $\text{Im}[G_F]$, decays with the distance away from the lightcone of this point. As evident from Figure \ref{regf}, the magnitude of the real part, $\text{Re}[G_F]$, is either close to zero outside of the lightcone of this point, or $\frac{1}{4}$ inside its lightcone. Thus the imaginary part of the propagator tells us the distance of the second point from the lightcone of the first point, while the real part indicates whether or not the second point is inside or outside the lightcone of the first (i.e., it indicates the causal structure). Figures \ref{imgf} and \ref{regf} therefore illustrate the intuition that the Feynman propagator  effectively provides a measure of the distance (or metric) between spacetime events of the causal set. The fact that knowledge of the Feynman propagator is to know metric distances helps explain why knowledge of the Feynman propagator is to know the spacetime manifold, i.e., in this case the causal set. 

It is not yet known whether an extension of this relation to the case of $4d$ causal sets is possible. Equation \eqref{gf2} continues to hold, but the general relation between $G_R$ and $C$ in $4d$ is not yet known. Therefore, given $G_F$ we can still deduce $G_R$ but we would not in general know how to deduce the causal matrix $C$ from it in $4d$.

\section{Towards Lorentzian Spectral Geometry}
As the Lorentzian counterpart to the Laplacian is the d'Alembertian, it is a natural operator to consider for spectral geometry on a causal set. 
For causal sets, it is known that d'Alembertians that have the correct continuum limit are nontrivially related to Green's functions that have the correct continuum limit (i.e, they are generally not simply inverses of each other),  \cite{sj1}. 
D'Alembertians that are known to possess the correct continuum limit can be constructed at a point by summing over values of the field on a few layers of elements to the past of that point. In this direction, there is a class of non-local d'Alembertians defined on causal sets \cite{dal, sia, lisa}. We will focus on the original d'Alembertian, $B$,  introduced in \cite{dal}.

If we fix an element $x \in \mathcal{C}$, at which we would like to know the value of	 $\Box\phi$ in the causal set, this will be 
\be
B\phi(x)=\frac{4}{\ell^2}\left(-\frac{1}{2}\phi(x)+\left(\sum_1-2\sum_2+\sum_3\right)\phi(y)\right)
\label{box1},
\ee\\
where $\ell$ is the discreteness scale and we are summing over $y$. We have separated the elements that precede $x$ into \emph{layers} according to the number of intervening elements between them and $x$. The first layer consists of those $y$ which are linked to $x$ such that $y \prec\!\!*~ x$, the second layer consists of those $y \prec x$ with only a single element $z$ such that $y \prec\!\!*~ z \prec\!\!*~ x$, etc. The causal set prescription for	$\Box \phi(x)$, \eqref{box1}, is then to take a combination  of the first few layers, with alternating signs and suitable coefficients. The three sums $\sum$ in \eqref{box1} extend over the first three layers as just described. 

As a matrix at point $x$, $B$ is
\be
\frac{\ell^2}{4}B_{xy}=\begin{cases}
-1/2, & \text{for $x=y$}\\
1, -2, 1, & \text{for $n(x,y)$= $0, 1, 2,$ respectively, for $x\neq y$}\\
0 & \text{otherwise}
\end{cases}
\ee

where $n(x,y)$ is the cardinality of the order-interval $\langle y, x\rangle=\{z\in\mathcal{C}|y\prec z\prec x\}$, or the number of elements of $\mathcal{C}$ causally between $y$ and $x$.

$B$ is linear, retarded, and invariant under relabelling of the causal set
elements. It can also be applied to any causal set, with or without curvature.

 In the continuum limit ($\ell\rightarrow 0$) the average of $B$  over  all  sprinklings  on  a  spacetime reduces to the continuum d'Alembertian plus a
term proportional to the Ricci scalar curvature \cite{fay}:

\begin{equation}
    \displaystyle{\lim_{\ell\rightarrow 0}}\, \bar{B}\, \phi(x)=(\Box-\frac{1}{2}R(x))\, \phi(x).
\end{equation}
Interestingly, as we will now show, given $B$, one can reconstruct $L$ (or $C$) and therefore the entire causal set. To this end, we search for matrix elements of value $1$ in $\frac{\ell^2}{4}B$, which denote links as well as cardinality 2 intervals. Checking for other relations between the elements, we can then distinguish between the links and intervals of cardinality 2. Once we have all the links, we have the link matrix $L$ and the causal set.

We conclude that there is a one-to-one correspondence between the link matrices for each causal set and their respective d'Alembertian. $B$ uniquely determines $L$, and $L$ uniquely determines a causal set. Hence this is a promising direction in which to explore spectral geometry. There are four levels of spectral geometry we can consider: 1) Whether the spectrum of $B$ or some other related operator can be used to distinguish ``manifoldlike" causal sets (i.e. ones that can be embedded into a Lorentzian manifold) from non-manifoldlike causal sets, 2) Whether the spectrum of $B$ or some other operator related to  $L$ can be used to distinguish causal sets that are different sprinklings into the same spacetime manifold, 3) Whether the spectrum of $B$ or some other operator related to $L$ can be used as a measure of how ``close" two  causal sets sprinkled into the same spacetime manifold are to one another, and 4) Whether the spectrum of $B$ or some other operator related to $L$ can be used to distinguish causal sets sprinkled into spacetime manifolds of differing curvature. In the latter case, causal sets obtained by sprinklings into a particular spacetime manifold should all possess the same (or approximately the same) spectra. A fundamental conjecture of causal set theory (called the ``Hauptvermutung" \cite{haupt}) is that two very different manifolds could not approximate the same causal set. Likewise, we do not expect to get similar spectra arising from very different manifolds. In Section 6 we find numerical evidence that the spectra are weakly able to distinguish different sprinklings into the same spacetime and that they are strongly able to distinguish sprinklings into different spacetimes. We also find some cases where the spectra can be used to distinguish causal sets that can be embedded into $1+1$D Minkowski spacetime from those that cannot.

Now the d'Alembertian $B$ itself is not self-adjoint. It is lower triangular, and its spectrum consists of eigenvalues that are all $-\frac{1}{2}$. Thus the spectrum of $B$ itself is trivial and is not useful for spectral geometry. Let us, therefore, consider  operators such as $(B\pm B^\dagger)$ and $(L\pm L^\dagger)$. We find that the spectra of such operators do indeed carry large amounts of geometric information.

\begin{figure}[htb!]
 \centering
 \includegraphics{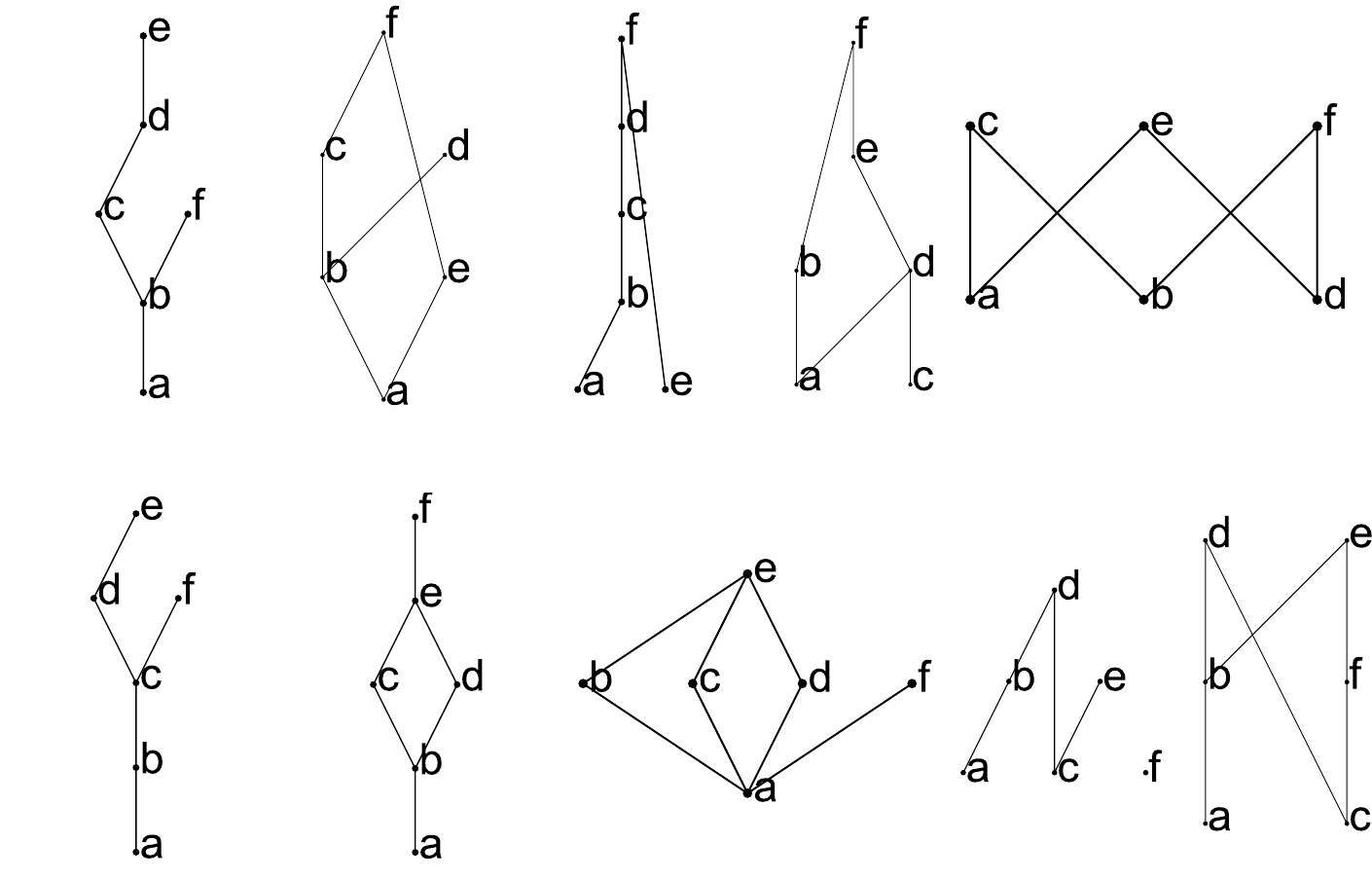}
 \caption{\label{hs}Hasse diagrams for a sample of 6 element causal sets. Lower elements precede higher elements and lines are drawn in for relations not implied by transitivity.}
\end{figure}

Concretely, for a small causal set, for example with 6 or 7 elements, we can enumerate all possible $L$'s and therefore all possible $B$'s. There are 318 possibilities with 6 elements and 2045 with 7 elements, including some that do not embed in $1+1D$ flat spacetime (such as the 6 element ``crown'' at the top right corner of Figure \ref{hs}). Hasse diagrams for a sample set of such 6 elements causal sets are shown in Figure \ref{hs}.  In a Hasse diagram, the causal set elements are represented as points, and lower elements precede higher elements. The lines are the relations not implied by transitivity. 

We will use the set of 6 and 7 element causal sets (or 6-orders and 7-orders) as one indication of the degree of uniqueness of the spectra of various causal set operators for a free massless scalar field in $2d$. Table 1 summarizes (in order of uniqueness) the results for the operators we considered: $G_F$, $G_F\pm G_F^\dagger$, $B\pm B^\dagger$, $i\Delta$, and $G_R+ G_R^\dagger$.

\begin{table}[h]
\begin{centering}
\begin{tabular}{| l | c | r |}
 \hline
    Operator & 
    6-orders: 318 total & 7-orders: 2045 total \\ \hline
    $i (B-B^\dagger)$ & 264 & 1709 \\ \hline
    $i\Delta$ & 210 & 1316 \\
    \hline
     $G_F$ & 201 & 1155 \\ \hline 
    $B+B^\dagger$ & 178 & 1013 \\ \hline
     $G_F+G_F^\dagger$ & 163 & 923 \\ \hline
    $G_R+G_R^\dagger$ & 162 & 921 \\
    \hline
     $i(G_F-G_F^\dagger)$ & 105 & 512 \\ \hline 
\end{tabular}
   \caption{\label{tab1} Approximate number of unique spectra for various causal set operators on 6- and 7-orders.}
\end{centering}
   \end{table}

Out of the operators we considered, the spectrum of $i(B-B^\dagger)$ does best\footnote{Since the degeneracies for different operators do not in general overlap with one another, a combination of the spectra of two or more operators could be used to distinguish between more causal sets.} at distinguishing between causal sets. It has the most number of unique spectra for the 6- and 7-orders. The degeneracy for some of the operators in Table 1 ($G_F$, $B+B^\dagger$,  $G_F+G_F^\dagger$, and $G_R+G_R^\dagger$) is never between spectra of causal sets that can be embedded into $1+1$D Minkowski spacetime and spectra of those that cannot (this is not the case for the remaining operators in Table 1, such as $i(B-B^\dagger)$ which has a small number of its degeneracies between manifoldlike and non-manifoldlike causal sets). Thus the spectrum of one or more of these operators might be a useful tool for distinguishing manifoldlike causal sets from non-manifoldlike ones.

We here only work with $1+1$D manifolds. In general, the asymptotics of the spectrum of the Laplacian is in one-to-one correspondence with the dimension of the manifold, according to Weyl's asymptotic formula \cite{chav}. If this asymptotic behavior can be translated into a property of also the d'Alembertian then one could envisage a constraint in the action that enforces a certain asymptotic behavior to obtain causal sets that can be embedded in  3+1 dimensional continuous spacetimes and energetically penalizes others. It should be very interesting to explore how natural such a constraint term may be. 

Another interesting observation that can be from the data in Table 1, is that the ratio of the number of different spectra of $i (B-B^{\dagger})$ to the number of different causal sets, is $0.830$ and $0.836$  for $n=6$  and $n=7$ respectively. This might suggest that the degeneracy could be removed for larger causal sets. Larger causal sets are necessary to test this conjecture, and we defer this investigation to future work.

We will next look more closely at the properties of the spectrum of $i(B-B^\dagger)$.

\section{The Spectrum of $i(B-B^\dagger)$}
\begin{figure}[htb!]
 \centering
 \includegraphics[width=.65\textwidth]{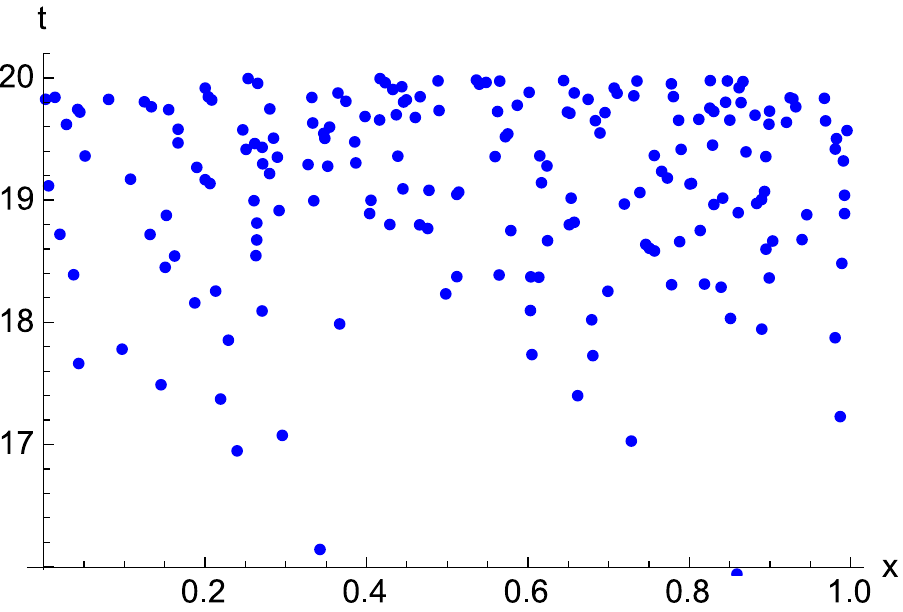}
 \caption{\label{conf}A causal set formed by sprinkling 200 elements into a finite interval in a conformally flat $2d$ spacetime with conformal factor $e^t$.}
\end{figure}

\begin{figure}[htb!]
 \centering
 \includegraphics[width=.65\textwidth]{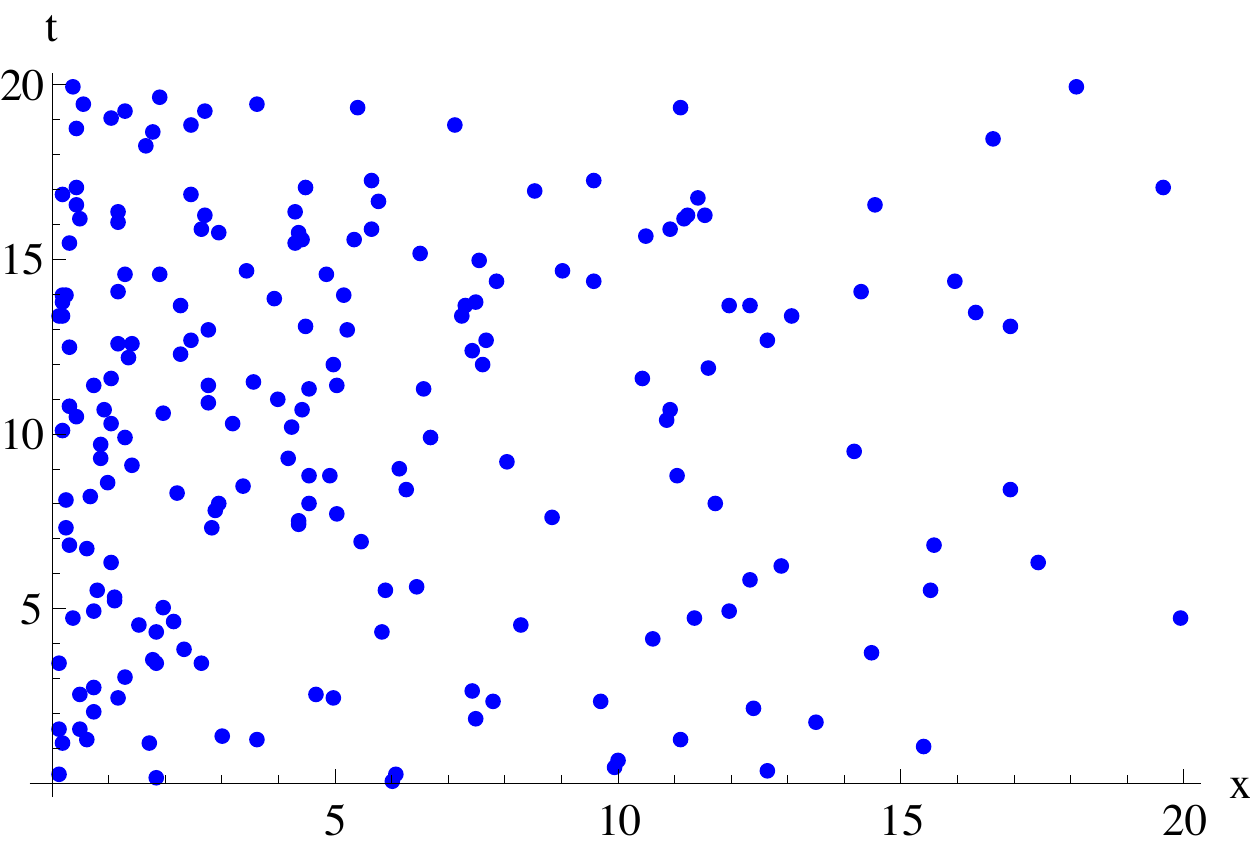}
 \caption{\label{inv}A causal set formed by sprinkling 200 elements into a finite interval in a conformally flat $2d$ spacetime with conformal factor $\frac{1}{1+x}$.}
\end{figure}

\begin{figure}[htb!]
 \centering
 \includegraphics[width=.65\textwidth]{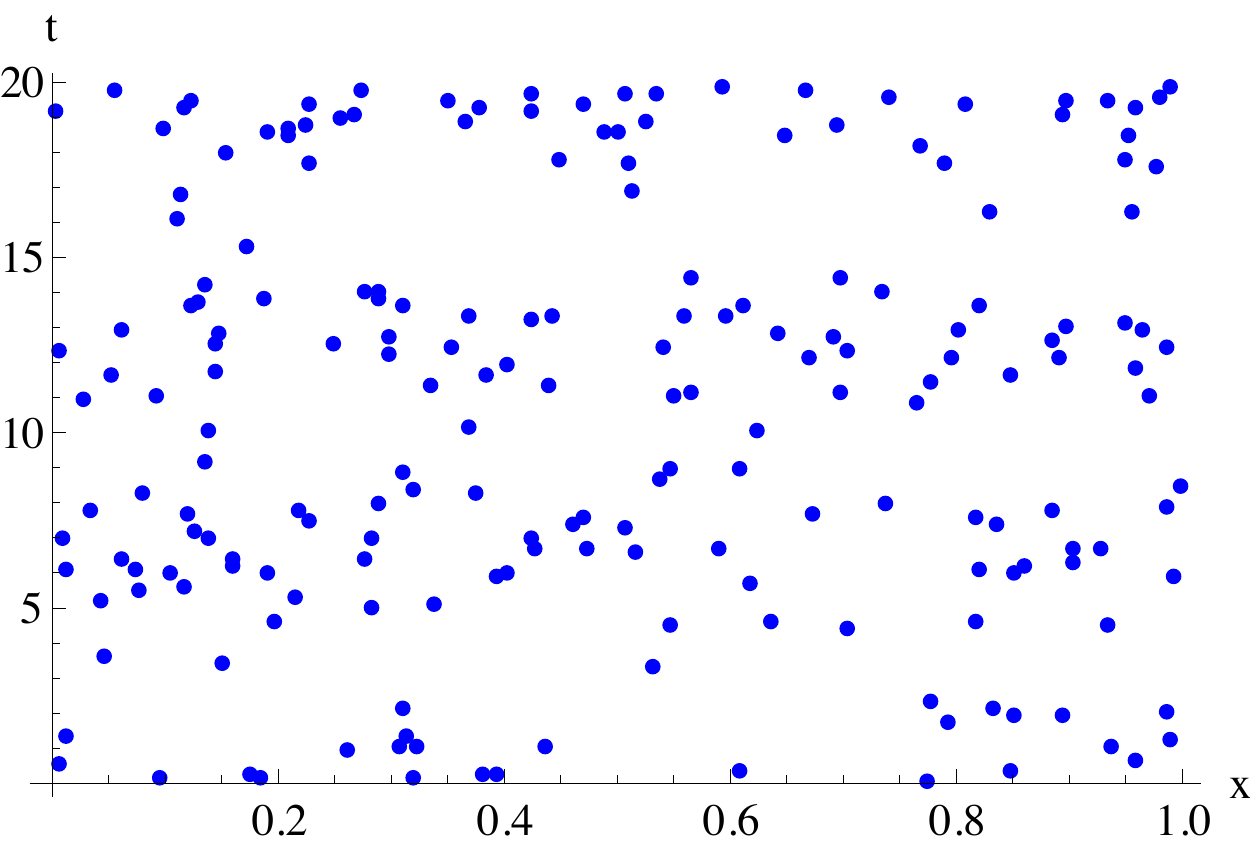}
 \caption{\label{cos}A causal set formed by sprinkling 200 elements into a finite interval in a conformally flat $2d$ spacetime with conformal factor $2+\text{cos}\, t$.}
\end{figure}

\begin{figure}[htb!]
\centering
 \includegraphics[width=.95\textwidth]{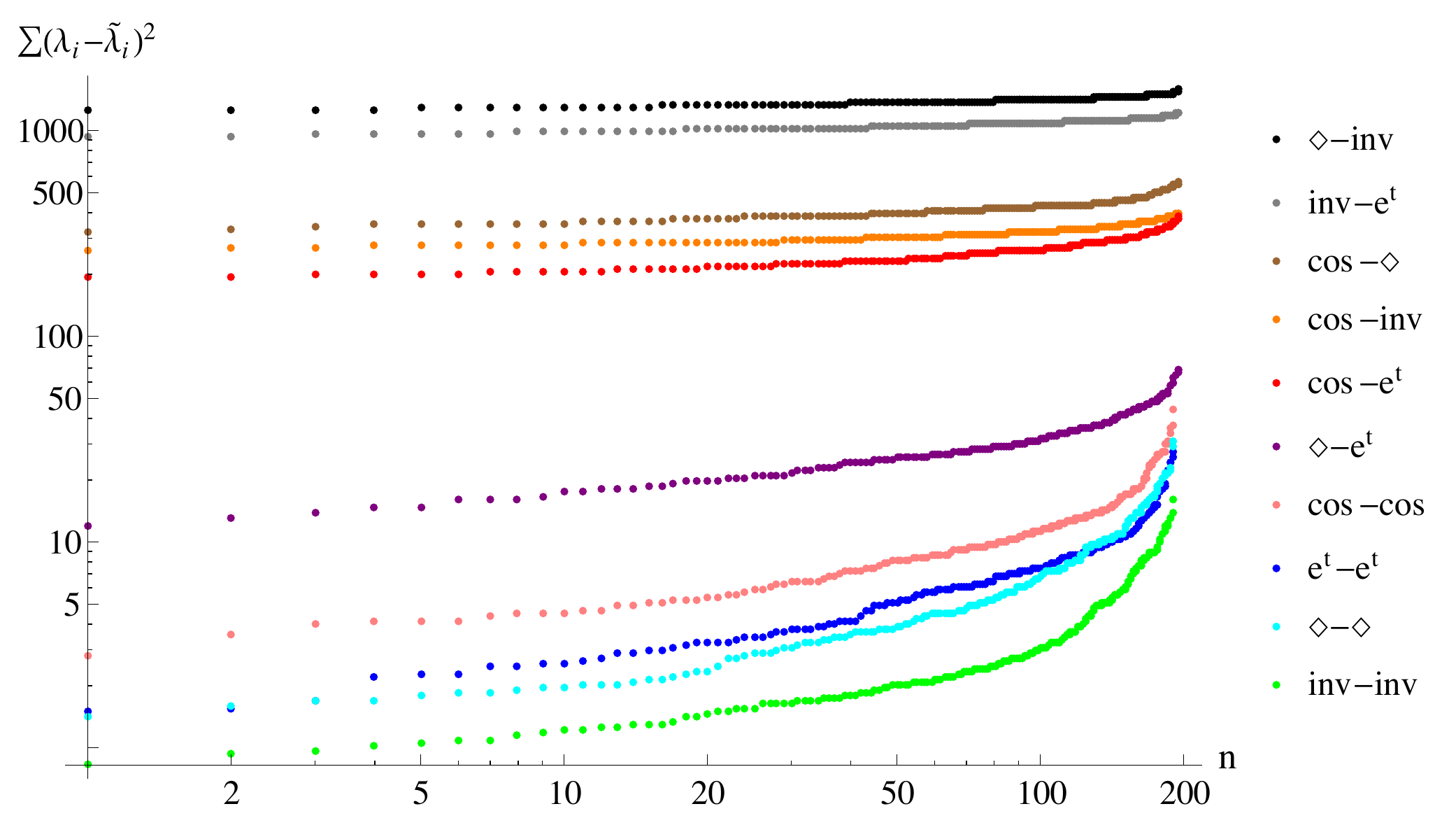}
 \caption{\label{comp}Spectral differences $\sum(\lambda_i-\tilde{\lambda}_i)^2$ for different sprinklings into the same and different manifolds. The horizontal axis labels the pair of sprinklings and the spectral differences are sorted in increasing order. $\Diamond$ corresponds to the spacetime of Figure \ref{2dcs}, $e^t$ corresponds to the spacetime of Figure \ref{conf}, \emph{inv} refers to Figure \ref{inv} and \emph{cos} refers to Figure \ref{cos}. The legend labels the curves from top to bottom.}
\end{figure}

Let us ask how the spectra of $i(B-B^\dagger)$ differ for sprinklings into different manifolds. We will consider 4 different manifolds: 1) A causal diamond in $2d$ Minkowski (Figure \ref{2dcs}, and $\Diamond$ in Figure \ref{comp}), 2) A patch of a $2d$ spacetime whose volume measure grows exponentially with time (Figure \ref{conf}, and $e^t$ in Figure \ref{comp}), 3) A patch of a $2d$ spacetime whose conformal factor is $\frac{1}{1+x}$ (Figure \ref{inv}, and $inv$ in Figure \ref{comp}), and 4) A patch of a $2d$ spacetime whose conformal factor is the oscillating function $2+cos\, t$ (Figure \ref{cos}, and $cos$ in Figure \ref{comp}).

We will work with 20 sprinklings of 200 elements into each of these spacetimes. Figure \ref{comp} shows the sum of the difference squared of the spectra of $i(B-B^\dagger)$ for pairs of sprinklings within each spacetime and across each pair of different spacetimes.  The numbers on the horizontal axis label  the pairs of spectra in the comparisons, and the spectral differences are sorted in increasing order for ease of comparison. The differences across two different spacetimes are clearly more pronounced. This means that if we are given two spectra and we find their difference to be large, we can say that there is a greater probability that they will correspond to different spacetimes than the same spacetime.

The second most unique operator in Table 1, $i\Delta$, also shows the same trend of spectral differences being larger across different spacetimes.

\section{Conclusions and Outlook}
We have shown that both the d'Alembertian $B$ as well as the Feynman Green's function $G_F$ contain the complete information about the causal set, at least for the 2-dimensional case. It would be very interesting to generalize these results to the 4-dimensional case. It can be shown that the $4d$ analogue of the $B$ we worked with in Section 4, introduced in \cite{dal}, also uniquely determines $L$. A challenge in exploring some of the other operators in Table 1 is that the general relation between $G_R$ and $L$ (or $C$) is not yet known in $4d$. Furthermore, while it is known that the $2d$ d'Alembertian $B$ that we have worked with leads to stable evolution, its $4d$ analogue has been shown to be unstable  \cite{sia}.  It is not known yet whether any of the $4d$ d'Alembertians in \cite{sia} lead to stable evolutions. We defer further investigation of the $4d$ case to future work.

That all geometric information is encoded in $G_F$ or $B$ means, in particular, that it should be possible to pursue the development of causal set kinematics, dynamics and quantization with the geometric degrees of freedom expressed in terms of  $B$ or $G_F$. The spectra of covariant operators such as d'Alembertians and correlators are geometric invariants, i.e., since they are labeling independent (analogously to diffeomorphism invariance). Therefore these spectral degrees of freedom could be easier to handle, for example, in a path integral where no modding out of spurious gauge (relabeling) degrees of freedom is required. Some ideas from \cite{gilk-old,hawk3,rovelli,sg2,ak1,ak2,tj,vanram} may prove useful in this direction.

In a number of approaches to quantum gravity, the phenomenon of the reduction of the spectral dimension on small scales  has been observed \cite{scar}. Work on the causal set approach to quantum gravity by Eichhorn and Mizera \cite{em} has indicated that, based on random walks and meeting probabilities,  the spectral dimension increases at small distances. In contrast, Belenchia et al. \cite{bel} computed the spectral dimension from the regularized Laplace transform of (causal set inspired) continuum non-local d'Alembertians using conventional heat kernel methods, and they find the usual spectral reduction to two dimensions in all cases. It should be interesting to pursue this question with the methods of the present paper, where we use spectral methods to capture the detailed shape of a spacetime, with the short distance structure expected to be encoded in the large eigenvalues.

Intuitively, the Feynman propagator, $G_F$, contains the geometric information because it contains the information about the metric: we know from the continuum theory that the Feynman propagator is a correlator of field fluctuations which decays with the distance. The Feynman propagator provides us, therefore, with a proxy for the metric distance - and to know the metric is to know the spacetime. 

The fact that also the Feynman propagator on a causal set contains the complete geometric information indicates that it too may serve as a proxy for the metric distances, in this case for causal sets. This suggests that it may be possible to use the Feynman propagator to calculate a new notion of a `metric on a causal set', analogously to how the metric can be calculated from the Feynman propagator in the continuum theory \cite{msa}. Our results in Figures \ref{imgf} and \ref{regf} illustrate this expectation.  

Further, we have shown numerical evidence that the spectra alone of correlators and d'Alembert operators already possess a large amount of geometric information about the underlying causal set: causal sets from sprinklings on the same manifold tend to have significantly closer spectra than causal sets from sprinklings on geometrically differing manifolds.  Intuitively, 
listening to the spectrum of the quantum noise on a causal set tends to tell about the causal set's geometric shape. The spectral distances of causal sets may therefore be useful for describing the dynamics or quantization of causal sets. 
$$$$
\bf Acknowledgements: \rm We thank the referees for their  constructive comments on an earlier version of the manuscript. AK acknowledges support in the form of a Discovery grant of the National Science and Engineering Research Council of Canada (NSERC). YY acknowledges support in part by Perimeter Institute for Theoretical Physics.
Research at Perimeter Institute is supported by the Government of Canada through
the Department of Innovation, Science and Economic Development Canada and by the
Province of Ontario through the Ministry of Research, Innovation and Science.

\end{document}